\documentclass[aps,prl,twocolumn,superscriptaddress]{revtex4-1}

\usepackage{array}
\newcolumntype{L}[1]{>{\raggedright\let\newline\\\arraybackslash\hspace{0pt}}m{#1}}
\newcolumntype{C}[1]{>{\centering\arraybackslash\vspace{-5pt}}p{#1}}
\newcolumntype{R}[1]{>{\raggedleft\let\newline\\\arraybackslash\hspace{0pt}}m{#1}}

\usepackage{amsmath}
\usepackage{mathtools} 
\usepackage{amssymb}
\usepackage{amsthm}
\usepackage{bm,upgreek} 
\usepackage{upgreek} 
\usepackage[table]{xcolor}
\usepackage{graphicx}
\usepackage{epstopdf}
\usepackage[colorlinks,linkcolor=blue,anchorcolor=blue,citecolor=blue,urlcolor=blue]{hyperref}
\usepackage{multirow}

\newcommand{\nn}{{\nonumber}}
\newcommand{\bea}{\begin{eqnarray}}
\newcommand{\eea}{\end{eqnarray}}
\newcommand{\ie}{\textit{i.e.{ }}}

\newcommand{\up}{\uparrow}
\newcommand{\dn}{\downarrow}

\newcommand{\md}{\mathrm{d}}
\newcommand{\me}{\mathrm{e}}

\newcommand{\nb}{NbSe$_2$/Nb$_3$Br$_8$/NbSe$_2$}

\begin{document}

\title{Symmetry Constraints on Direct-Current Josephson Diodes}

\author{Da Wang}  
\author{Qiang-Hua Wang} 
\affiliation{National Laboratory of Solid State Microstructures $\&$ School of Physics, Nanjing University, Nanjing 210093, China}
\affiliation{Collaborative Innovation Center of Advanced Microstructures, Nanjing University, Nanjing 210093, China}

\author{Congjun Wu} \email{wucongjun@westlake.edu.cn}
\affiliation{Department of Physics, School of Science, Westlake University, Hangzhou 310024, Zhejiang, China}
\affiliation{
Institute for Theoretical Sciences, Westlake University, Hangzhou 310024, Zhejiang, China}
\affiliation{Key Laboratory for Quantum Materials of Zhejiang Province, School of Science, Westlake University, Hangzhou 310024, Zhejiang, China}
\affiliation{Institute of Natural Sciences, Westlake Institute
for Advanced Study, Hangzhou 310024, Zhejiang, China}

\begin{abstract}
It is necessary to break both time-reversal and parity symmetries
to realize a Josephson, or superconducting, diode exhibiting
nonreciprocal critical direct-currents (DC).
In fact, these conditions are still insufficient.
The dependencies of the free energy on the phase difference across the junction
and the magnetic field are classified, exhibiting the current-reversion (JR), field-reversion, and field-current reversion conditions, respectively.
To exhibit the DC Josephson diode effect, all symmetries satisfying the
JR condition need to be broken.
The relations of critical currents with respect to the magnetic field are classified into five classes,
including three exhibiting the diode effect.
These symmetry considerations are applied to concrete examples.
Our work reveals that the DC Josephson diode effect is a natural
consequence of the JR symmetry breaking, hence, providing a guiding
principle to understand or design a DC Josephson diode.
\end{abstract}
\maketitle

\emph{Introduction}.
The semiconductor diode plays a fundamental role in modern electronics.
A Josephson diode effect with nonreciprocal supercurrent was firstly
proposed by Hu, one of the author, and Dai \cite{Hu2007} in junctions
between hole- and electron-doped superconductors close to a Mott insulator which works as the depletion region.
When the junction is forward-biased by an electric field $E$, the depletion region shrinks exhibiting
an alternating-current (AC) Josephson effect with a large critical current $J_c$,
while the depletion region expands under a reverse bias which
effectively shut down the Josephson junction, giving $J_c(E)\ne J_c(-E)$.
On the other hand, recently, a nontrivial diode effect has been observed recently in
many experiments in the direct-current (DC) Josephson effect, exhibiting nonreciprocal critical current $|J_{c+}| \ne |J_{c-}|$ ($\pm$ labels the forward and
backward directions)  \cite{exp:Ando2020, exp:Baumgartner2021, exp:Bauriedl2021, exp:Diez-merida2021, exp:Farrar2021, exp:Idzuchi2021, exp:Lin2021, exp:Lyu2021, exp:Miyasaka2021, exp:Pal2021, exp:Shin2021, exp:Wu2021, exp:Hou2022, exp:Golod2022, exp:Narita2022, exp:Gupta2022, exp:Turini2022}.
These progresses have triggered a great deal of theoretical studies on the supercurrent diode effect in superconductors \cite{th:Misaki2021, th:He2021, th:Yuan2021, th:Daido2021, th:Halterman2021, th:Zhang2021, th:Zinkl2021, th:Jiang2021, th:Scammell2021, th:Davydova2022, th:Zhai2022, th:Karabassov2022, th:Souto2022, th:Legg2022, th:Tanaka2022, th:Jiang2022}.

We focus on the DC Josephson diode effect below.
The AC Josephson diode effect is essentially a non-equilibrium phenomenon,
while the DC one is an equilibrium property.
An important question is to figure out the conditions for such an effect.
Researchers have recognized that all of time-reversal ($\+T$) and parity symmetries [including inversion ($I$) and mirror reflection ($M$)] have to be broken,
since any of them protects $|J_{c+}|=|J_{c-}|$ \cite{th:He2021, th:Yuan2021, th:Jiang2021, th:Davydova2022}.
However, one may still find various systems with all the above symmetries
broken but still do not exhibit the diode effect.
One known example is the Josephson junction connecting two different
materials (breaking $I$ and $M$) with magnetic impurities in between (breaking $\+T$), where the magnetic scattering
could give rise to a $\pi$-junction \cite{Buzdin2005} but does not exhibit the diode effect.
Other examples will also be given in the context below.
Therefore, except the easily distinguished $\+T$, $I$ and $M$ breaking conditions,
there still exist additional constraints to realize a DC Josephson, or superconducting, diode.
A unified picture, if exists, is highly desired for future
studies in this field.

In this article, we examine what symmetries forbidding the DC Josephson diode effect,
and conversely, all such symmetries have to be broken to realize such
an effect.
According to even-or-odd dependence of the free energy with respect to
the supercurrent $\0J$ and magnetic field $\0B$ reflections, there are three relations
including current-reversion (JR), field-reversion (BR), and field-current
reversion (BJR) ones.
Any unitary or anti-unitary symmetry leading to the JR relation forbids
the diode effect.
The relations of critical current $J_{c\pm}$ with respect to the magnetic field $\mathbf{B}$
are classified into five classes, including three exhibiting the diode effect.
Model Hamiltonians are constructed with the magnetic field and
spin-orbit coupling (SOC) to examine the above relations, which reveals
symmetry constraints more stringent than the previous time-reversal
and parity ones.

\begin{table*}
\begin{tabular}{|C{0.9cm}|C{1.6cm}|C{2.7cm}|C{2.7cm}|C{2.8cm}|C{2.8cm}|C{2.8cm}|}
\hline
                                                      &                   & \multicolumn{3}{c|}{DC diode effect}   & \multicolumn{2}{c|}{absence of DC diode effect}              \\ \hline
                                                      &        & BJR (type-I)   & BR (type-II)    & none (type-III)      & JR\&BR\&BJR & JR     \\ \hline
                                                      &  \vspace{30pt} sketches of $J_{c\pm}(B)$   & $\includegraphics[width=.9\linewidth]{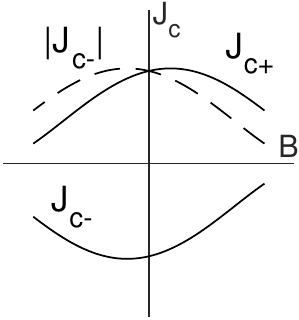}$     & $\includegraphics[width=.9\linewidth]{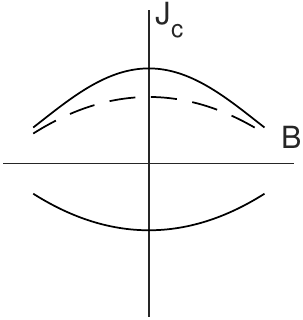}$         & $\includegraphics[width=.9\linewidth]{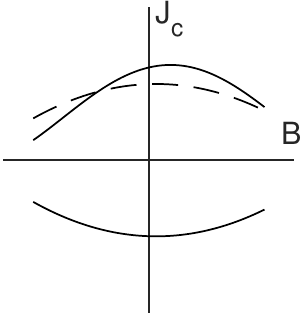}$  &   $\includegraphics[width=.9\linewidth]{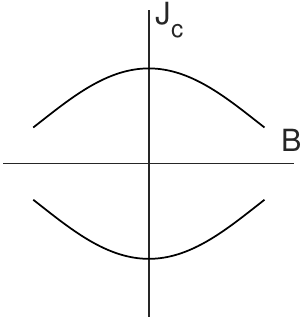}$      & $\includegraphics[width=.9\linewidth]{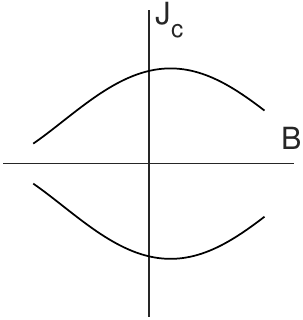}$   \\ \hline
\multirow{2}{1cm}{\vspace{5pt}{ } $\0J\parallel\hat{\0x}$, $\0B\parallel\hat{\0x}$} & \vspace{5pt}broken symmetries     & $I$, $M_{x,y,z}$, $I\+T$, $\+TM_{x,y,z}$, etc.  & $\+T$, $I$, $M_x$, $C_{2y,2z}$, $\+TM_{y,z}$, $\+TC_{2x}$, etc.      & $\+T$, $I$, $M_{x,y,z}$, $C_{2y,2z}$,  $\+TM_{x,y,z}$, $\+TC_{2x}$, $I\+T$, etc.   & \vspace{10pt} {none}        & $\+T$, $M_{y,z}$, $C_{2y,2z}$, $I\+T$, $\+TC_{2x}$, $\+TM_x$, etc.  \\ \cline{2-7}
                                                      & satisfied symmetries  & $\+T$, $C_{2y,2z}$, $\+TC_{2x}$, etc. & $M_{y,z}$, $I\+T$, $\+TM_x$, etc.       & \vspace{5pt} none  & $(I,\+T)$, $(I,C_{2y,2z})$, $(\+T,M_x)$, etc.      & $I$, $M_x$, $\+TM_{y,z}$, etc.  \\ \hline
\multirow{2}{1cm}{\vspace{5pt}{ } $\0J\parallel\hat{\0x}$, $\0B\parallel\hat{\0y}$} & \vspace{5pt} broken symmetries        & $I$, $M_z$, $C_{2x,2y}$, $I\+T$, $\+TM_z$, $\+TC_{2x,2y}$, etc.  & $I$, $\+T$, $M_x$, $C_{2y,2z}$, $\+TM_{y,z}$, $\+TC_{2x}$, etc.      & $I$, $\+T$, $M_{x,z}$, $C_{2x,2y,2z}$, $\+TM_{y,z}$, $\+TC_{2x,2y}$, $I\+T$, etc.  &\vspace{10pt}{none}        & $\+T$, $M_{x,z}$, $C_{2x,2z}$, $I\+T$, $\+TM_y$, $\+TC_{2y}$, etc.  \\ \cline{2-7}
                                                      & satisfied symmetries  & $\+T$, $M_x$, $C_{2z}$, $\+TM_y$, etc. & $M_z$, $C_{2x}$, $I\+T$, $\+TC_{2y}$, etc.      &  \vspace{5pt}none  & $(I,\+T)$, $(I,M_x)$, $(I,C_{2y,2z})$, etc.   & $I$, $C_{2y}$, $\+TM_z$, $\+TC_{2x}$, etc.   \\ \hline
\end{tabular}
\caption{Usual time-reversal and spatial symmetry constraints for the five classes of DC Josephson junctions.
Without lose of generality, we assume the current $\0J$ along the $\hat{\0x}$-direction and the magnetic field $\0B$ along $\hat{\0x}$- or $\hat{\0y}$-directions.
For each case, we list the symmetries to be {\it all} broken, and, to be satisfied {\it with at least one}.}
\label{table:diode}
\end{table*}

\emph{Current and field reversion conditions}.
We consider a Josephson junction with a superconducting phase difference
$\Delta \phi$ across the junction.
When the orbital effect of the magnetic field $\0B$ needs to be considered,
$\Delta \phi$ should be replaced by the gauge invariant version
$\Delta \tilde \phi=\Delta \phi -(2e/\hbar)\int\mathbf{A}\cdot \md\bm{\ell}$,
where $\mathbf{A}$ is the vector potential to be integrated along the path across the junction.
In addition, $\mathbf{B}$
also breaks $\+T$ in the spin channel and hence manifests in the free energy $F$ as well.
In the most generic case, the free energy $F$ could
exhibit no symmetry with respect to  $\mathbf{B}$
and $\Delta\tilde \phi$.
Nevertheless, the symmetries of an experimental system may lead
to the following relations,
\bea
\label{eq:JR}
\text{JR}:\quad F(\mathbf{B} ,\Delta\tilde \phi)&=&
F(\mathbf{B},-\Delta\tilde\phi+\theta), \\
\label{eq:BR}
\text{BR}:\quad F(\mathbf{B},\Delta\tilde \phi)&=&
F(-\mathbf{B},\Delta\tilde\phi+\theta), \\
\label{eq:BJR}
\text{BJR}:\quad F(\mathbf{B},\Delta\tilde \phi)&=&
F(-\mathbf{B},-\Delta\tilde\phi+\theta),
\eea
where $\theta$ is a constant global phase.
Eqs.~\ref{eq:JR}, \ref{eq:BR}, and
\ref{eq:BJR} are termed as current-reversion (JR),
magnetic-field-reversion (BR), and magnetic field and current
simultaneously reversion (BJR) relations, respectively.

The above free energy conditions give rise to different symmetry relations of
the critical supercurrents.
The supercurrent is defined as $J=(2e/\hbar)\partial F/\partial \Delta \tilde \phi$,
and the critical supercurrents $J_{c\pm}$ are
maximal values in the forward and backward directions, respectively.
The following convention is employed that $J_{c+}>0$ and $J_{c-}<0$.
Eq.~\ref{eq:JR} leads to
\bea
\text{JR}:\quad
J_{c+}(\mathbf{B}) &=& -J_{c-}(\mathbf{B}),
\label{eq:J_JR}
\eea
which protects the absence of the diode effect.
In contrast, Eq.~\ref{eq:BR} and Eq.~\ref{eq:BJR} give rise to
\bea
\text{BR}:\quad
J_{c\pm}(\mathbf{B})&=&J_{c\pm}(-\mathbf{B}),
\label{eq:J_BR} \\
\text{BJR}:\quad
J_{c+}(\mathbf{B})&=&-J_{c-}(-\mathbf{B}).
\label{eq:J_BJR}
\eea
In both cases, the Josephson diode effect appears if the JR relation is violated.

If any two of the above three conditions are satisfied, then the third
one is automatically true.
Hence, there exist five types: all conditions,
one of the JR, BJR, and BR conditions, and none,
are satisfied, as shown in Table~\ref{table:diode}.
The former two situations show the absence of the diode effect,
while the latter three exhibit it, labeled by type-I, II, III,
respectively.
For the type-I diode, the curves of $J_{c\pm}$ versus $\0B$ are central symmetric, i.e. satisfying Eq.~\ref{eq:J_BJR}, which is widely observed in many experiments \cite{exp:Ando2020,exp:Baumgartner2021,exp:Bauriedl2021,exp:Farrar2021,exp:Diez-merida2021,exp:Idzuchi2021,exp:Lin2021,exp:Miyasaka2021,exp:Pal2021,exp:Shin2021,exp:Hou2022,exp:Golod2022,exp:Narita2022,exp:Gupta2022,exp:Turini2022}.
As for the type-II diode, the BR condition protects
$J_{c\pm}(\mathbf{B})=J_{c\pm}(-\mathbf{B})$, which
is reported in the InAs-junctions under a background magnetic field \cite{exp:Baumgartner2021} and the \nb-junctions \cite{exp:Wu2021} under zero external magnetic field.

We next analyze concrete symmetries leading to the above free energy conditions.
As an example, a unitary symmetry $U$, or, an anti-unitary symmetry $U\+K$ ($\+K$ the complex conjugate) leading to
the BJR relation can be expressed as
\bea
U^\dag H(\mathbf{B},\Delta \tilde \phi) U = H(-\mathbf{B},-\tilde \Delta \phi
+\theta), \label{eq:4symB}
\label{eq:unitary}
\eea
or,
\bea
U^\dag H^*(\mathbf{B},\Delta \tilde \phi) U = H(-\mathbf{B},-\tilde \Delta \phi
+\theta).
\label{eq:anti-u}
\eea
Either one leads to Eq.~\ref{eq:BJR}
and in turn Eq.~\ref{eq:J_BJR}.
The symmetry conditions leading to Eqs.~\ref{eq:JR} and
\ref{eq:J_JR}, and to Eqs.~\ref{eq:BR} and \ref{eq:J_BR} can be constructed similarly.

For type-I diode effect, we consider two typical configurations of the relative directions for $\0B$ and $\0J$: parallel and perpendicular.
When they are parallel, e.g. $\mathbf{B}\parallel \mathbf{J}\parallel
\mathbf{\hat x}$,
the 2-fold rotation symmetries along the $\hat{\0y}$ and $\hat{\0z}$-axes ($C_{2y}, C_{2z}$)
satisfy Eq.~\ref{eq:unitary}, while $\cal T$ and
the combination of $\cal T$ with the 2-fold rotation around the $\hat{\0x}$-axis,
${\cal T} C_{2x}$, satisfy Eq.~\ref{eq:anti-u}.
In either case, we arrive at the BJR condition of Eq.~\ref{eq:BJR}, which leads to the central symmetry of the curves of $J_{c\pm}$ versus $\0B$ described by Eq.~\ref{eq:J_BJR}.
On the other hand, the curves of $J_{c\pm}$ violate the reflection symmetries,
indicating
the Hamiltonian should break all of the following symmetries leading to BR and JR conditions,
including inversion $I$, mirror reflections with respect to
the $yz$, $xz$, and $xy$-planes ($M_i$ with $i=x,y,z$), and the combined
symmetries of ${\cal T} I$, ${\cal T} M_{i}$ ($i=x,y,z$), etc.
If the field and the current are perpendicular, say, $\0J\parallel\hat{\0x}$ and $\mathbf{B}\parallel\hat{\0y}$,
similar conclusions can be drawn.
Any one of the following symmetries, $\cal T$, $M_x$, $C_{2z}$,
${\cal T} M_y$, is sufficient to lead to the BJR condition.
Conversely, all the following symmetries leading to BR and JR should be broken,
$I$, $M_z$, $C_{2x}$, $C_{2y}$, ${\cal T}I$,
${\cal T}M_z$, ${\cal T}M_z$, and ${\cal T} C_{2y}$, etc.

Similar analysis can be straightforwardly applied to other
situations including the type-II and type-III diode effect, and also for the other two cases exhibiting no diode effect, as summarized in Table~\ref{table:diode}.
The symmetry patterns are much richer than previous results in literature,
and they provide a guidance to design DC Josephson, or superconducting,
diodes in future studies.

\begin{figure}
\includegraphics[width=0.3\textwidth]{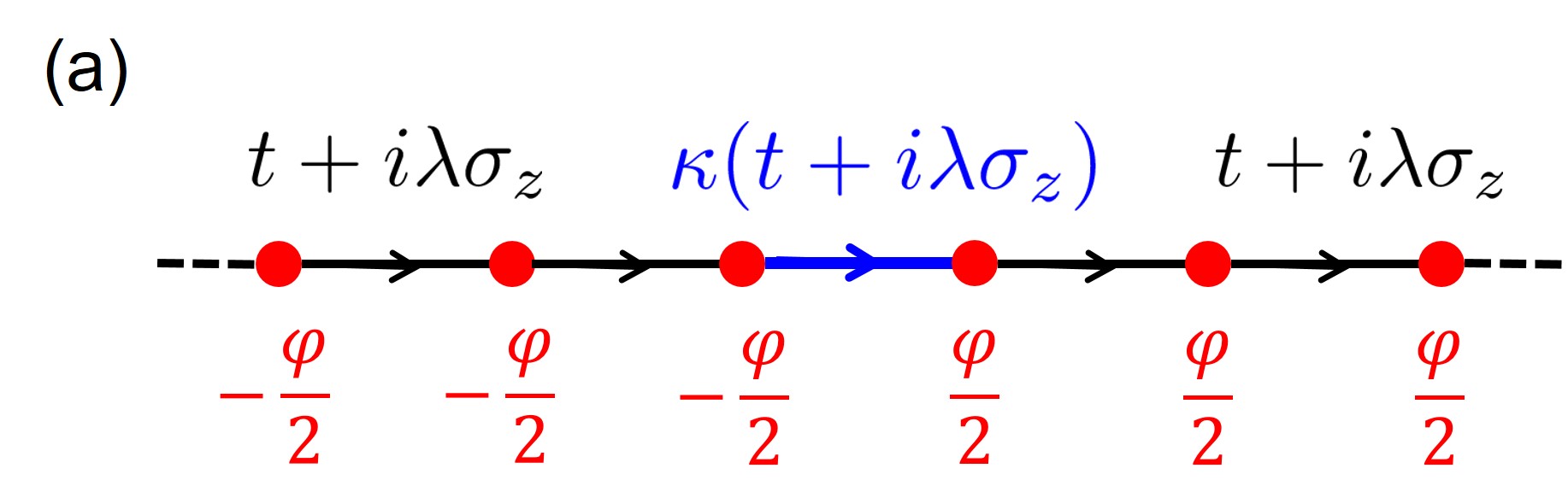} \\
\vspace{5pt}
\includegraphics[width=0.4\textwidth]{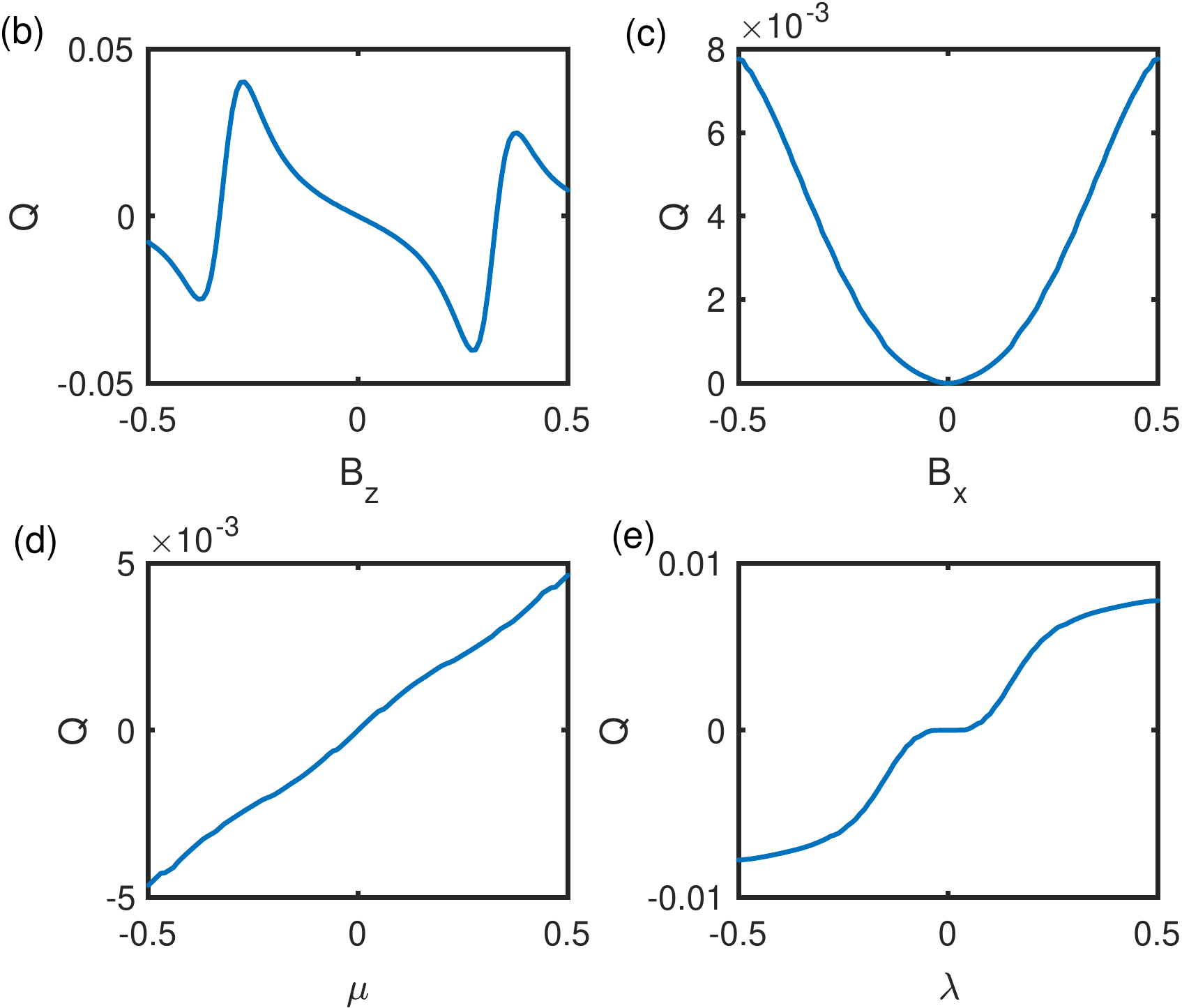}
\caption{(a) Scheme of the superconducting chain along $\hat{\0x}$-direction.
The hopping $t$ and SOC strength $\lambda$ are reduced by a factor $\kappa$ on the interface (blue) bond. The superconducting phases on the two leads are $\pm\varphi/2$. From (b) to (e), the nonreciprocal factor $Q$ defined in Eq.~\ref{eq:Q} is plotted with varying $B_z$, $B_x$, $\mu$ and $\lambda$, respectively, while keeping the others fixed at $B_z=B_x=\lambda=0.5$, $\mu=1$, $\Delta=0.2$ and $\kappa=0.4$.}
\label{fig:1d}
\end{figure}

\emph{1D model and the DC Josephson diode effect}.
We proceed to consider concrete models to verify the above
symmetry conditions of the critical supercurrents.
We first consider a superconducting chain along the
$\hat{\0x}$-axis as shown in Fig.~\ref{fig:1d}(a).
The model Hamiltonian of the Bogoliubov-de Gennes (BdG) mean-field
theory reads
\bea
H_{\rm 1D}&=&\sum_i \left[c_{i}^\dag \left(t_i+ i\lambda_i\sigma_z \right)
c_{i+1} + h.c. \right] \nn\\
&+& \sum_i c_i^\dag\left(-\mu-\0B\cdot\bm{\sigma}\right)c_i
+ \left(\Delta_i c_{i\uparrow}^\dag c_{i\downarrow}^\dag + h.c.\right),
\label{eq:ham1d}
\eea
where $c_i$ is a two-component spinor, $t_i=t$ is the nearest neighbor hopping inside two leads taken
as the energy unit, and $t_i=\kappa t$ on the interface bond,
$\mu$ is the chemical potential, $\Delta_i$ is the spin-singlet pairing with phases $\pm\frac{\varphi}{2}$ on two sides,
respectively.
The Zeeman field $\mathbf{B}$ lies in the $xz$-plane.
The SOC quantization axis lies along the $\hat{\0z}$-axis, with the strength $\lambda_i=\lambda$ inside two leads and $\kappa\lambda$ on the interface bond.
By calculating the free energy $F$ as a function of $\Delta\phi=\varphi$,
which is just the ground state energy at $T=0$K for simplicity,
the Josephson current is obtained as $J(\varphi)= (2e/\hbar)
\partial_\varphi F$, whose maximal/minimal values by varying $\varphi$ give
$J_{c\pm}$, respectively.

Such a system breaks the inversion by the SOC term and breaks time-reversal by the Zeeman term, respectively.
Hence, naively one would expect a DC Josephson diode effect.
However, such a diode effect
only appears when all the quantities of
$\lambda$, $B_z$, $B_x$, and $\mu$ are simultaneously nonzero.
In Fig.~\ref{fig:1d}(b)-(e), the nonreciprocal factor $Q$, defined as
\bea \label{eq:Q}
Q=\frac{|J_{c+}|-|J_{c-}|}{|J_{c+}|+|J_{c-}|},
\eea
is plotted with varying each one of the parameters $B_z$, $B_x$, $\mu$ and
$\lambda$, while fixing the others at nonzero values.

The above results show that as long as one of the parameters becomes zero,
the DC Josephson diode effect vanishes because at least one symmetry
leads to the JR condition of Eq.~\ref{eq:JR}.
(I) If $\lambda=0$, the
inversion $I$ is a unitary symmetry satisfying the
JR condition, \ie leaving the Hamiltonian invariant but reversing the supercurrent.
(II) If $B_z=0$, the JR relation is satisfied due to
the mirror reflection symmetry $M_x$.
After switching on $B_z$, the JR condition is violated,
and $M_x$ reflects $B_z$ satisfying the BJR condition, giving the type-I diode effect in Table~\ref{table:diode}.

The situations of $B_x=0$ or $\mu=0$  involve new symmetries not shown in Table~\ref{table:diode}.
(III) For $B_x=0$, we first define a spin-twist operation
$U_{tw}$ as a position-dependent spin rotation
\bea
U_{tw}=\prod_{i} U(i) ,
\eea
with $U(i)=e^{i\frac{\sigma_z(i)}{2}(i-1)\eta}$ acting on site-$i$ and $\eta= \arctan(\lambda_z/t)$.
The spin twist leaves the $B_z$ term unchanged, eliminates the SOC term, and transform it into the hopping term by replacing
$t$ with $\sqrt{t^2+\lambda^2}$ in Eq.~\ref{eq:ham1d}.
Since the pairing is spin-singlet, it is not changed
by this spin rotation.
Then a combined operation $U=U_{tw} I U_{tw}^\dagger$
leaves the Hamiltonian invariant except switching the current direction.
For nonzero $B_x$, the $\pi$-rotation in the spin space $R_z(\pi)=\me^{i\frac{\pi}{2}\sigma_z}$ brings $B_x$ to $-B_x$ without reflecting the current, satisfying the BR condition and hence the diode effect belongs to type-II in table~\ref{table:diode}.
(IV) For $\mu=0$, the particle-hole transformation $c_{i}\to (-1)^i\sigma_yc_i^\dag$,
brings $H(\Delta_i)$ to $H(-\Delta_i^*)$, hence, reversing the supercurrent.
Actually, the particle-hole symmetry only applies to a bipartite lattice,
but it can be a good approximation when the band is
near half-filling in general.
From this 1D toy model, we have seen the possibilities of JR symmetries
beyond the usual time-reversal and spatial (inversion, mirror
and $\pi$-rotation) ones.

\begin{figure}
\centering
\includegraphics[width=0.5\textwidth]{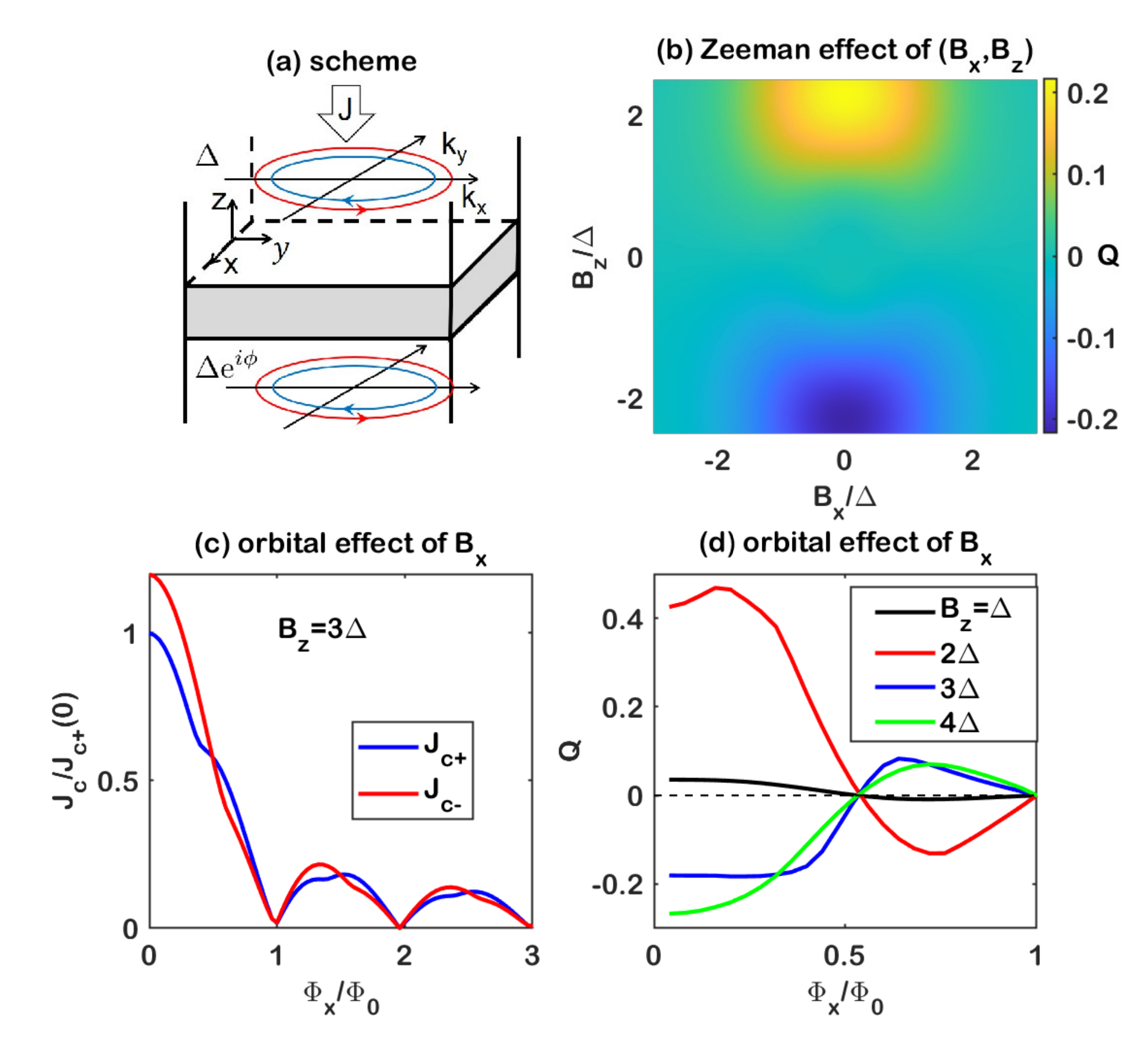}
\caption{(a) Scheme of the Josephson junction with the supercurrent along the $\hat{\0z}$-direction. The Fermi surfaces are split due to the Rashba SOC and the spin directions are indicated by small arrows with different colors. (b) shows the nonreciprocal factor $Q$ versus $B_x$ and $B_z$ with only Zeeman effect considered. In (c) and (d), the orbital effect of $B_x$ is taken into account, and the results of $J_{c\pm}$ and modified $Q$ (see definition in the main text) with respect to $\Phi_x$ are given, respectively.
In numerical calculations, we choose the parameters: $\Delta=0.2$, $\mu=3$, $\lambda_R=0.5$, $t_z=\lambda_z=0.4$.}
\label{fig:tunnelz}
\end{figure}

\emph{2D toy model with out-of-plane current}.
We next consider a bilayer toy model connected by a narrow junction
schematically shown in Fig.~\ref{fig:tunnelz}(a).
The tunneling direction is along the $\hat{\0z}$-direction.
The BdG Hamiltonian reads
\begin{align}
H_{\rm 2D}=&\sum_{n=t,b,\0k} \left\{c_{n\0k}^\dag\left[\varepsilon_\0k\sigma_0 + \lambda_R(k_x\sigma_y-k_y\sigma_x) -\0B\cdot\bm{\sigma} \right]c_{n\0k} \right. \nn\\
&\left.+\left(c_{n\0k\up}^\dag \Delta_n c_{n-\0k\dn}^\dag + h.c.\right)\right\} \nn\\
&+\sum_\0k c_{t\0k}^\dag (t_z+ i\lambda_z\sigma_z ) c_{b\0k} +  h.c.,
\label{eq:Ham2d}
\end{align}
where $t$/$b$ refers to the top/bottom layer,
$\0k=(k_x,k_y)$ is the in-plane momentum,
$t_z$ the tunneling matrix element.
$\lambda_R$ is the Rashba SOC,
and $\lambda_z$ is another SOC coupling to $\sigma_z$.
The Zeeman field $\mathbf{B}$ is assumed to
lie in the $xz$-plane.
The spin-singlet pairings $\Delta_t=\Delta$ and $\Delta_b= \Delta\me^{i\varphi}$, and then $\Delta \phi=\varphi$.
The band dispersion is simply chosen as $\varepsilon_\0k= -2t(\cos k_x +\cos k_y)
-\mu$ with $t$ taken as the energy unit.

The nonreciprocal factor $Q$ is plotted on the $(B_x,B_z)$ plane as shown in Fig.~\ref{fig:tunnelz}(b).
$B_x$ itself cannot lead to a
nonzero $Q$ unless a nonzero $B_z$ exists.
Again, all the quantities of $B_z$, $\lambda_z$ and $\lambda_R$ need to
be nonzero for the appearance of the diode effect.
(I) If $B_z=0$, then the combined symmetry $\+TC_{2z}$, leads to the JR condition and protects $Q=0$.
For nonzero $B_z$, since $\+TC_{2z}$ reflects $B_z$ and current simultaneously, satisfying the BJR condition,
the curves of $J_{c\pm}$ vs $B_z$ belongs to type-I in table~\ref{table:diode}.
On the other hand, the $B_x$-dependence belongs to type-II in table~\ref{table:diode}, since $C_{2z}$ brings $B_x$ to $-B_x$ satisfying the BR condition.
(II) If $\lambda_z=0$, the combined symmetry $\+T M_x$
leads to the JR condition and hence protects $Q=0$.
(III) If $\lambda_R=0$, the symmetry satisfying the JR condition is a little subtle.
We first perform a spin twist $U_{tw}=
\me^{-i\frac{\eta}{4}{\sigma_{z,t}}}\me^{i\frac{\eta}{4}{\sigma_{z,b}}}$ with
$\eta= \arctan(\lambda_z/t_z)$ to eliminate the $\lambda_z$-SOC term, which transforms $B_x\sigma_x$ to $B_x(\cos\frac{\eta}{2}\sigma_x \pm \sin\frac{\eta}{2}\sigma_y)$ ($\pm$ for top/bottom).
Then we perform a $\frac{\pi}{2}$-rotation around $\sigma_x$, i.e. $R_x(\frac{\pi}{2})=\me^{i\frac{\pi}{4}\sigma_x}$, to obtain $(B_x\cos\frac{\eta}{2}\sigma_x\pm B_z\sigma_y\mp B_x\sin\frac{\eta}{2}\sigma_z)$, which is invariant under the current-reflecting operation $C_{2x}$. At last, the above operations are inversely applied to recover the original Hamiltonian except $\Delta_n\to\Delta_n^*$. Put them together, we obtain the combined JR symmetry $U=U_{tw}R_x(\frac{\pi}{2})C_{2x}R_x(-\frac{\pi}{2})U_{tw}^\dag$ which protects $Q=0$.

We tentatively compare our results with
the experiment on the \nb-junctions \cite{exp:Wu2021}.
The DC Josephson diode effect is observed at zero external magnetic
field and is suppressed by the in-plane field $B_x$, showing the type-II behavior in Table~\ref{table:diode}.
According to the symmetry principle, the time-reversal must already be broken at $B_x=0$.
It is then natural to conjecture the existence of a spontaneous ferromagnetic
moment (still labeled by $B_z$ for our convenience) \cite{FM:Zhu2016, FM:Wickramaratne2020}.
(This assumption may be at odds with the absence of longitudinal magnetoresistivity hysteresis with the $\hat{\0z}$-directional external magnetic field \cite{exp:Wu2021}, which deserves further studies.)
Another possibility for the $\+T$-breaking is the pairing itself breaks time-reversal like in $s+id$ or $p+ip$ superconductors \cite{th:Zinkl2021}.

In our model, the $\lambda_z$-SOC term is necessary to cause the diode effect.
This term requires $M_z$ breaking, which is indeed possible since the interface Nb$_3$Br$_8$ does break $M_z$ \cite{FM:Pasco2019}.
As a comparison, when the interface is replaced by few-layer graphene preserving $M_z$ symmetry, the $\lambda_z$-term is forbidden, leading to absence of the (external) field-free diode effect \cite{exp:Wu2021}.
In Fig.~\ref{fig:tunnelz}(b), we find $Q$ is evenly suppressed by $B_x$, {\it qualitatively} similar to the experiment, but the required field strength (of order $\Delta/\mu_B\sim 10$T) is much larger than the experimental value $\sim10$mT \cite{exp:Wu2021}.
This small field strength indicates that $B_x$ should couple to $J_z$ mainly through the orbital effect rather than the Zeeman effect.
In this regard, we choose the vector potential $A_z=yB_x$
entering into $t_z$ and $\lambda_z$ through the Peierls phase $\me^{i(\Phi_x/\Phi_0)(y/L_y)}$ ($\Phi_0$ the magnetic flux quantum and $L_y$ the magnetic unit cell).
Within the quasiclassical picture,
the supercurrent $J(\varphi,\Phi_x,y)$ is integrated over $y$
to obtain $J(\varphi,\Phi_x)$ and then $J_{c\pm}(\Phi_x)$.
In Fig.~\ref{fig:tunnelz}(c) and (d), $J_{c\pm}$ and the modified nonreciprocal factor $Q=[J_{c+}(\Phi_x)+J_{c-}(\Phi_x)]/ [J_{c+}(0)-J_{c-}(0)]$ are plotted versus $\Phi_x$, displaying a modulated nonreciprocal Fraunhofer pattern. The feature at small field with weak $\Phi_x$-dependence is similar to the experimental results \cite{exp:Wu2021}.

\emph{Summary}.
In summary, we have specified three types of symmetry conditions, i.e. JR, BR and BJR, based on which the relations of $J_{c\pm}$ vs $\0B$ are classified into five classes, including three exhibiting the DC Josephson diode effect.
These symmetry constraints provide a unified picture to understand or design a DC Josephson diode in future studies.



\emph{Acknowledgement}.
D.W. thanks H. Wu, X. Xi for helpful discussions and also thanks S.-J. Zhang for early collaborations in this project. This work is supported by National Natural Science Foundation of China through the Grants Nos. 11874205, 11729402, 12174317, 12234016, 12274205 and 11574134.

\bibliography{scDiode}
\end{document}